# CONNECTING THE DOTS IN NUTRITIONAL REHABILITATION: A QUALITATIVE STUDY ON ICT AND COMMUNITY BASED CARE

Deepa Austin, International Institute of Information Technology, Bangalore, deepa.austin@iiitb.ac.in

Amit Prakash, International Institute of Information Technology, Bangalore, amitprakash@iiitb.ac.in

**Abstract:** 'Fragmentation in care' continuum is often considered as a shortcoming of Health system whereas, 'Integration of care' is widely acclaimed as a viable solution to fragmentation. In last two decades, Information and communication technologies (ICTs), by virtue of their ability to integrate information for action, has been extensively used in addressing many public health problems like malnutrition. Tackling the public health challenge of malnutrition demands attention to interconnectedness and interactions between multiple systems. In this paper, using a case study of an ICT application used by community workers for malnutrition management in Karnataka, we argue that lack of recognition of interconnectedness and interactions among stakeholders and context can pose a challenge to integration of care. ICTs can be key enablers to overcome fragmentation, provided it recognizes the inherent complexities of malnutrition and its management. We argue that for an effective ICT enabled integration of Severe Acute Malnutrition (SAM) management, a thorough understanding of perspectives of multiple stakeholders together with rich picture of the contextual dynamics should not be ignored at design and implementation phase.

## 1. INTRODUCTION

Information and communication technologies (ICTs) are considered as key catalyst for change towards integrating information for action(Walsham, 2020). In the last two decades, ICTs has had a profound influence on health system, by connecting the otherwise fragmented system into a more integrated one(Srivastava et al., 2015). 'Fragmentation in care' often cited as one of the shortcomings of current health system, has been defined as "systemic misalignment of incentives, or lack of coordination, that spawns inefficient allocation of resources or harm to patients, adversely impacting quality, cost, and outcomes"(Enthoven, 2009).The advancements in science and technology has led to greater specialization and a "silo approach" in health system, weakening its inherent integrated nature. While fragmentation in care arises from lack of interaction among stakeholders and systems, 'integration of care' on the other hand acknowledges this and appreciates the inherent complex nature of health systems(Braa et al., 2017). However, the distorted understanding of dependencies and interactions among multiple systems embedded in their respective contexts poses a challenge to integrated care continuum. This research is focused on understanding the usage of ICT applications in attending to challenges in integrating the health services for managing Severe Acute Malnutrition (SAM) in Karnataka, India.

Public Health challenge of Malnutrition has a complex nature, characterized by interconnectedness and interactions among many systems. Management of malnutrition demands attention to the 'Whole' calling for a perspectival change from a fragmented to a more holistic approach(SPRING, 2015). A recent study has shown that COVID-19 crisis has aggravated the scenario resulting in an estimated increase of 14.3% in child wasting globally (Headey et al., 2020). It is therefore estimated that with 47 million children globally (under 5 years of age) affected by wasting before pandemic, an additional 6·7 million children would be added during the first 12 months of the pandemic accounting to an estimated 10,000 additional child deaths per month during this same period. According to this study, Sub-Saharan Africa and South Asia could be the worst affected region. Condition is no better in India, as NFHS-5 data(2019-20) showed that state of Karnataka has 35.4 % of stunting, 19.5 % of wasting, 32.9% of underweight and 8.4 % of severe wasting among children under 5 years of age (*National Family Health Survey (NFHS-5)*, n.d.).India is yet to assimilate the immediate, intermediate and long term effects of this pandemic on malnutrition status, however, building up resilience in all spheres of intervention strategies remain crucial. The interwoven nature of people, resources and organizations that are characteristic to health systems should possess resilience to cope up with such unprecedented situations. Under such circumstances, it becomes more relevant to strengthen the call for a multi-sectoral approach initiated by Indian National Nutrition Mission to ensure a holistic continuum of care with an ICT enabled Nutrition Information System for tracking and monitoring the entire process(*POSHAN Abhiyaan*, 2018).

A participatory action research (PAR) coupled with an agile methodology to design and develop software applications for nutritional rehabilitation was taken up at E-Health Research Centre of IIIT-Bangalore from September 2018, in collaboration with a technology team (CSTEP) and Nutritional Rehabilitation Centre (NRC) at a medical institution in state of Karnataka. The overall objective of this project is to 'Track and Manage Severe Acute Malnourishment' (SAM) through community-based care centers [Anganwadi(AWC) and Primary Health Centre(PHC)] and facility-based care centers attached to hospitals(NRC). The initial phase of this research (Sept 2018 till Dec 2019) studied two processes–referral process of SAM children from AWC to PHC/NRC (using ethnography) and follow-up process from NRC (using in-depth interviews). The findings are used in the design and development of an integrated digitized platform for SAM management connecting community and facility centers. By June 2020, a web-based platform for NRCs across the state of Karnataka was successfully launched and is functioning at its best. For this paper, we have put together the findings from the referral processes. The findings of this study contribute to the knowledge that ICTs can be key enablers to overcome fragmentation, provided it recognizes the inherent complexities of malnutrition and its management. We argue that for an effective ICT enabled integration of SAM management, a thorough understanding of perspectives of multiple agents together with rich picture of the contextual dynamics should not be ignored at design and implementation phase.

## 2. THE COMPLEXITY OF HEALTH SYSTEM

Before we delve into the problem of malnutrition, it is essential to understand the complexity science of health system in general. World Health Organization defines a health system as a "system that consists of all organizations, people and actions whose primary intent is to promote, restore or maintain health". It aims at improving health and health equity in ways that are responsive, financially fair making the best use of available resources (de Savigny et al., 2009). The components of health system interact nonlinearly over multiple scales, rendering the system more dynamic nature with unpredicted outcomes. This is in contrast to mechanical systems with well-defined boundaries, in which component parts interact linearly and manipulation of each part is expected to produce a predictable output(Plsek & Greenhalgh, 2001). Kuipers et al. conceptualized the health system complexity under three main headings: Medical Complexity encompassing the spectrum of illness to wellness, Situational complexity that takes into consideration personal factors (age, gender, race, Socioeconomic status, Education, cultural factors), physical, social and other environmental factors influencing health(Pim Kuipers, Elizabeth Kendall, et al., 2011). The complex interplay of medical and situational complexities leads to geometric progression of system complexities. System complexities include service fragmentation, critical decision making, uncoordinated practice management, ethical dilemmas, commercialisation and commoditisation, improper budgeting, unparalleled prioritization, over impacting technological determinism, inappropriate health policies and regulatory framework and most importantly, lack of humanitarian values(Greenhalgh & Papoutsi, 2018; Safford et al., 2007).

Complexity in health systems is attributed to interconnectedness and interactions between multiple agents acting across systems (Braa et al., 2017). Considering the intertwined nature of health system, it is appropriate to consider it as a complex adaptive system in which a collection of individual agents (human and non-human) with freedom to act in ways that are not always totally predictable, and whose actions are constitutively entangled so that its influence on each other is mutual (Orlikowski, 2007; Plsek & Greenhalgh, 2001). Such thinking in systems perspective can offer a thorough understanding of multiple interacting agents together with their contexts in a comprehensive way, thus helping to anticipate synergies and mitigate negative emergent behaviors.

## 3. THE PROBLEM OF FRAGMENTATION AND ITS SOLUTION

World Health Organization recognizes fragmented care as one of the five common shortcomings of current healthcare delivery system. (World health report, 2008). Fragmentation in care continuum fails to acknowledge the interconnectedness and the resultant interactions in complex health system.

Stange in his interactive series on 'the problems in fragmentation' narrates a situation in which a patient is moved from specialist to specialist, each of them studying in-depth the organ in which s(he) is an expert, passing through the latest of the tests, prescribing the best medicine, ultimately feeling desolate and fragmented ending with fatigue. He then quotes "Healing requires relationships—relationships which leads to trust, hope, and a sense of being known. Increasingly our health system delivers commodities that can be sold, bought, quantified, and incentivized. While the whole—whole people, whole systems, and whole communities—gets worse"(Stange, 2009). Hence, due acknowledgement of these interconnectedness and relationships becomes indispensable.

The problems of fragmentation in care identified by Stange includes 'inefficiency' in rendering wholesome efforts to prioritization of needs using available resources, 'ineffectiveness' in achieving a comprehensive, universal, equitable and affordable healthcare service to all. This paves way to health 'inequality' and inequity favoring the inverse care law by Hart(Hart, 1971).When thick strands of relationships woven on trust and care are replaced by thin strands of profit and gain, healthcare systems struggle to survive. Short term goals increase 'Commoditisation' and 'Commercialisation' leading to 'deprofessionalization' and 'depersonalization' further weakening our health system. The end result would be 'despair and discordance' with the system.

A viable solution to fragmented care is 'Integrated care' as it has the potential to facilitate interconnectedness. Integrated care drives the health system from a disease-oriented to a more person-oriented service sector, integrating the right mix of innovations and social care, coordinating and administering it through strategic management approaches. Though there is no unifying definition for Integrated care, it is often used synonymously to terms like coordinated care and seamless care(Kodner & Spreeuwenberg, 2002; World health organization, 2016). Such integrations can be brought about at organizational, functional, service or clinical level.

## 4. NUTRITIONAL REHABILITATION SERVICES

Government of India (2018) targets to reduce stunting, under-nutrition, anaemia and reduce low birth weight by 2%, 2%, 3% and 2% per annum respectively by year 2022.Nutrition rehabilitation requires a life cycle approach rooting in early preventive action, supplementary nutrition, sanitation and education on health and hygiene. *At community level*, greater emphasis on valuing, recognizing and enhancing contribution of Anganwadi workers(AWWs), and their empowerment is envisaged in POSHAN Abhiyaan. Under the flagship of Integrated Child Development Services by Ministry of Women and Child Development (MWCD), Screening for malnutrition and Supplementary Nutrition programme is being implemented through Anganwadi Centres(AWCs) for addressing under-nutrition in pregnant and lactating women and under-6 children. The Nutritional Rehabilitation Centres(NRC) are *facility-based care* units that can be fitted into the chronic care model of integrated care where, SAM children below six years are admitted with their mothers for treatment, stabilization and rehabilitation. Timely interventions and continued monitoring can prevent deaths due to SAM(Aprameya et al., 2015).This requires a continuum of care streamed through NRCs for SAM children with medical complications and community-based care for those without medical complications.

## 5.    THE INTEGRATED MODEL OF SAM CARE

AWWs carry out regular growth monitoring at AWCs. If the child falls low on nutritional indicators consecutively for 3 months, they are referred to nearest PHC. For patients with no associated co-morbidities, treatment and medication are given from PHC. In case, In-patient care is required then they are referred to NRC. After the condition improves, the patient is sent back to the community. In ideal case, follow-up should occur until the child is out of malnutrition cycle. The Figure 1. depicts the integrated model of care in Karnataka.

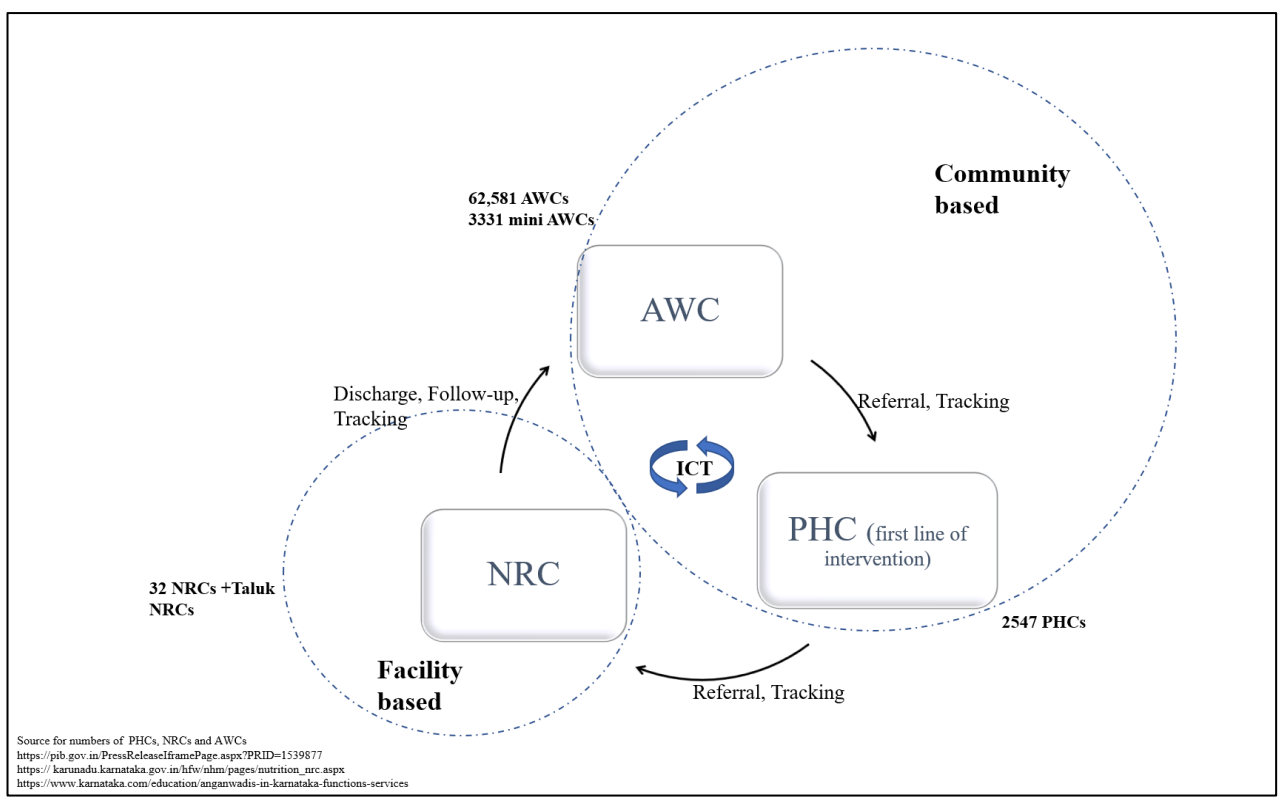

**Figure 1. Nodes for Integrated SAM care in Karnataka**

## 6.    RESEARCH DESIGN

This research is part of a larger action research project trying to address the problem of SAM through an ICT enabled continuum of prevention, cure, rehabilitation and follow-up by connecting and integrating various service locations like AWCs, PHCs and NRCs. Although PAR is the methodology for entire project, ethnography was used in this study for two reasons- firstly, it was part of academic requirement for the researcher (a female doctoral student at IIIT-B), secondly, a multi-sited and comparative ethnography was the best choice of methodology towards the *objective* of understanding daily work practices of AWWs towards addressing malnutrition as well as understanding the perceived and the actual role of ICT in facilitating and integrating the care continuum[here,  a mobile based application used by AWWs - SNEHA (Solution for Nutrition and Effective Health Access)]. The field observations were conducted from March 2019 to May 2019.

The findings from this study were later incorporated in the framework on 'challenges in integrating nodes of care delivery' and were considered during design, development and implementation of the entire project. The Institutional Review Board approval for the study was granted by IIIT- B. Prior to commencement of the study, permission was sought through email from respective authorities.

Informed oral consent: The interviews were conducted in person by the researcher under oral consent, in preferred language of participants (with the help of an interpreter for those participants not eloquent in English/ Hindi). The purpose of the study was explained in a manner that they could understand. Audio and video recordings were done with permission. Participants were informed that they were free to withdraw from interviews or from study whenever they wished. In this study, no one refused.

Sampling: A purposive sampling was done and four AWCs were selected as study settings. One anganwadi where SNEHA application was first piloted in Karnataka (May 2018) was selected for two reasons-provided sufficient time to understand the functioning of SNEHA and had large beneficiary numbers that allowed us to understand challenges in data handling. Three other AWCs working in conventional way without using SNEHA but within the vicinity of the first chosen AWC allowed us to have better comparison as the contextual dynamics almost remained the same. Observations and interviews were carried out until data saturation achieved.

| Data collection techniques | |
|---|---|
| **Techniques** | **Remarks** |
| Immersive participant observation (12 weeks) | The daily work practices of AWW teachers (4 in number aged between 35 to 50 years) and helpers (4 in number aged between 30 to 40 years) were observed by the researcher through the study period. Other participants observed included beneficiaries of ICDS, Auxiliary Nurse Midwife (ANM) and Accredited Social Health Activist (ASHA), Male health workers and beneficiaries of Anti-natal and post-natal services, who visited the AWCs |
| Informal interviews (15 to 30 mins, 20 interviews) | wherever required for better clarifications relating to work practices |
| Intensive interviewing using interview guide (90 to 120 mins, 7 interviews) | Of AWW teachers and helpers to understand their roles towards malnutrition and their perceptions on SNEHA application. Interviewing of technology developer to understand the challenges during/after roll out |
| Archival enquiry | walked through manual registers and SNEHA application to understand the recording and maintenances practices. |
| Physical traces | of growth charts, posters were assessed to understand the work practices. |

**Table 1. Data collection techniques used**

A field journal was maintained to note down observations and reflexivity of the researcher on a daily basis. All the interviews were transcribed in English along with notes of observation. Major findings were orally discussed with the participants. The field journals were sifted systematically through close reading by researcher, manual open coding, memos and thereafter focused coding and integrative memos to arrive at final themes from the data. For this paper, we have used only those findings relevant to challenges in integration.

# 7. SNEHA application

SNEHA is an android application for AWWs launched by Centre for Study of Science, Technology and Policy (CSTEP) under the expert guidance of Department of Women and Child Development (DWCD), Karnataka. The application currently works on Samsung J5 Series Android 7.1.1 provided by DWCD whereas mobile SIM card is given by CSTEP. The front end is pure native Android with the middleware Java programming language. Current back end is Postgres, but considering the scalability in future, discussions are on to move with Microsoft SQL server. Data base is currently local to CSTEP, but discussions are on to move it to cloud owned by Government of India.

| User interface of SNEHA mobile application (at the time of the study) | Objectives of SNEHA mobile application |
|---|---|
| **Child services** include daily attendance, daily nutrition, height, weight, Middle upper arm circumference (MUAC), immunization details and other derived parameters for identifying stunting, wasting and severe acute malnourishment (SAM) which are colour coded for easy identification.<br>**Mother service** includes Mathru-poorna, Thaayi card details, RCH id, Growth status, ANC/PNC, incentives (Mathru-Vandana), tracking options.<br>**Special services** include census survey details, any special national program related details like pulse polio<br>**Anganwadi corner** details out priorities to be carried out day to day as well as monthly basis and a trouble ticketing system, where an AWW can voice out their concerns | 1. Maternal and child health management (connecting the health registries of mother and child and integrative referral services made accessible across the country)<br>2. Multi-sectoral process integration (integrate the activities of DWCD and Ministry of health and family welfare)<br>3. Multi-system data entry (digital solutions to multiple registers maintained by an AWW)<br>4. Inclusive solutions for AWW which intends to ease out their daily activities and voice out their concerns |

**Table 2. Domains in user interface of SNEHA and Objectives of SNEHA application**

# 8. MAJOR FINDINGS

## 8.1 The social proximity and acceptance of AWWs in the community they serve and their role in mediating the activities towards health in general.

The researcher observed the daily work routine of AWWs like imparting informal education through songs, acts and games, their interactions with children, record keeping, use of digital devices, inventory maintenance, growth monitoring, nutritional supplementation for children and pregnant ladies, information transfer on government schemes to adult beneficiaries, conducting awareness sessions etc. Major findings from researcher's observations were

1. Being selected from the local community and designated as 'community worker', an AWW has close social ties with the community they serve.
2. This helped them to gain better acceptance among the community in terms of understanding their medical needs and appropriately directing them to avail medical services.
3. An AWW could bridge the gap between the culture of community and culture of medicine by mediating and exercising information transfer in terms of health check-ups/referrals/follow-ups and nutrition awareness sessions.

4. It was observed that reminder calls and oral alerts were given to those who missed out routine immunization services, growth monitoring services and ante- and post-natal care. The data from an AWW would serve as an alert for health department to plan for preventive strategies.
5. By virtue of their social proximity and community acceptance, an AWW could offer the best foster care for a SAM child.

## 8.2 Diverse roles to be exercised

An AWW maintains around 11 prescribed paper-based registers centered around their job responsibilities.

One of the AWW's who never used SNEHA remarked that her time was mostly wasted in finding the right register to enter the details. Some of these entries were often postponed and carry forwarded to home, so that she could give more attention to children at AWC. At home, she could do the entries with minimal disturbance and limited errors. But, while doing so, there were chances that some details get missed out. Issues of duplication of data entry for immunization was observed as both AWWs and ANM recorded and maintained the same details in two different registers- for AWC and PHC. Another concern was repeated entry of details especially, when a child/ pregnant lady reports back after a long period of absence (generally happens when family migrates to other places of work). Thus, most of the man hours seemed to be spent on manual data entry and back tracking the records. Also, it was observed that a good amount of time was spent on explaining the documents required for various government schemes like Thaayi card, Aadhar card etc. and their follow up with adult beneficiaries. Information exchange was also carried out over telephonic conversations. There were challenges as to inadequate space to maintain the registers safe, issues of power outage and poor network connectivity for making phone calls.

When asked about the use of SNEHA to a regular user, she replied *"Overall, the app is good and easy to use. I can choose the language according to my comfortability, many are using Kannada. But to me, I like to use English. That way, I can learn some new words in English. When I type, I feel error is less and that makes me confident. I just need to enter the height and weight. It shows some colored graph. I don't have to go after the papers (referring to growth chart)."* It is interesting to note that while most of HCI researchers advocate the use of different languages in the interface for better accessibility, it may not be the only value attached to it. According to this AWW, SNEHA application not only has made her data entry easy, but also instilled confidence in her by reducing the errors in recording data and helped her to use a language that she wanted to learn.

Here is an excerpt of an interview with technology developer of SNEHA, *"99% of the AWW's are happy with the application. They say it is easy to use. Within seconds they get access to the data pertaining to their AWC. One of the greatest challenges that we faced was the internet connectivity. Later, based on the signal strength, the best network service provider was chosen. We are planning to go offline too."*

While SNEHA has eased out some of the daily tasks like attendance, back tracking of data, yet some challenges were observed by the researcher pertaining to the context in which the application was to function like infrastructural inadequacies and need for digital skilling. SNEHA was generally handled by Anganwadi teacher. In case the teacher goes on leave, daily activities may not be updated due to two reasons- mobile phone with application was not made available with helpers and lack of knowledge of data entry by helpers.

## 8.3 Activities towards SAM assessment- Anthropometric measurements

Growth monitoring is an essential activity towards nutritional status assessment of a child. The field visits gave the researcher a grim picture of inaccuracies in anthropometric measurements (weight and height measurement). A spring hanging scale (SALTER) that can weigh up to 25 kg and is graduated by 0.1kg (100g) increments was used to weigh children. Out of four AWCs, three of them lacked calibrated weighing scales, had torn weighing pant bags in which children were placed to be weighed. Two AWCs had torn height scales on their walls. None of the AWCs had infantometer and electronic weighing scales to measure weight of infants. These challenge the very purpose of active screening of children below 2 years.

AWWs who were not using the mobile application did not record the main diagnostic criteria of SAM- the weight/ height ratio. When asked, an AWW replied *"This is relatively new measurement, we are used to weight-for-age criteria. Our supervisors have not asked us to use the new standards. Wt/Ht ratio requires some calculation (dividing Wt/Ht to find 'Z' scores)."* It is important to make a note here, that three AWCs without mobile application did not report any case of SAM, whereas AWC using SNEHA reported 4 children falling under SAM category. It can be inferred that SNEHA has been instrumental in making use of right indicator towards right diagnosis.

When asked to demonstrate, all four AWWs showed height measurement process by making the child stand straight. Two of them made the child lean against the wall. Others did by making them stand wherever they were. When the researcher took the measurement in the WHO prescribed way, by making the child lean against the wall with head, shoulder, buttocks and heels touching the wall, their measurements differed by 2 to 4 cm. This would alter the weight/ height ratio (Wt/Ht), a strong indicator of SAM. Wrong measurements led to wrong entry and diagnostic errors in identifying SAM children and adversely affecting the health of such children. Again, this brings us to the understanding that mere digitization cannot rule out wrong diagnosis.

### 8.4  Referral of SAM

When enquired about the referral processes to PHC, an AWW replied, "..*if the child remains underweight for three months, then we refer. We take the paper records of last three months (ht, wt recorded and food given) along with us to PHC. If a child doesn't need hospitalization, then a counselling session is given to parents on diet related matters by PHC doctor. Medicines are delivered through AWC, and parents collect them. Earlier they had to wait for Rs.2000/- to be sanctioned from DWCD and then buy medicine. The entire process used to take 3 to 4 months' time. Till then, we would manage the child with food alone. Now medicines are given by the government, still it is a slow process. Takes about three to four weeks' time. Some vitamins and deworming tablets are provided from PHC. Till we get medicines, we manage by feeding the child with healthy food like egg and milk. If hospitalization is essential, then doctor refers them to nearest NRC."*

When enquired, *"Has any of your beneficiary gone to NRC for treatment?"*, an AWW replied, *"Yes, one child had gone to NRC. She used to get flu and fever often. She was referred to NRC from PHC. But, don't know why? they sent her back. Even the PHC doctor doesn't know the reason. No instructions given to me regarding her health. She is still weak. I feed her egg and milk daily."* This excerpt clearly shows the current state of disconnect among the AWC, PHC and NRC further leading to fragmentation in services.

### 8.5 Other issues

Children belonging to APL (above poverty line) category hardly visited the AWCs. In one of the centres, it was told that out of 10 APL children, only 3 had enrolled in AWC. Due to time constraints, this study has not gone in-depth on such issues.

Based on the observations and interviews, a summary of challenges in integration is given below (Table 3.)

| | | Nodes of integrated care delivery | | |
|---|---|---|---|---|
| | | AWC | PHC | NRC |
| Challenges in integration | process challenges | gaps in referral, follow ups, Duplication of activities inadequate active screening, APL category not recorded, issues with migration | gaps in referral, follow ups, Duplication of activities, no data base of APL | Inadequate community referral, gaps in screening, monitoring and follow ups |
| | Lack of resources(Physical and technical) | Physical space constraints, time constraints, lack of digital devices to connect with PHC/NRC and integrate data(for those not using SNEHA), network connectivity issues, lack of calibrated devices for Ht, Wt measurements, financial constraints | Difficulty in connecting with AWC/NRC to monitor and provide feedback, time constraints, network connectivity issues | Difficulty in connecting with AWC/PHC to monitor and provide feedback for the discharged patients |
| | individual level inadequacies | inadequate trainings resulting in errors in measurement, Incorrect awareness about significance of nutritional indicators, Need for digital skilling | inability to carry out follow ups(attributed to lack of resources) | inability to carry out follow ups(attributed to lack of resources) |
| | Policy gaps | no uniformity in the standards used for assessment of malnutrition, Changing indicators | no uniformity in the standards used for assessment of malnutrition, Changing indicators, inadequate follow-up policies | no uniformity in the standards used for assessment of malnutrition, Changing indicators, inadequate follow up policies |
| | Regulatory challenges | disconnected means of policy enforcement | disconnected means of policy enforcement | disconnected means of policy enforcement |
| | Functional challenges | cumbersome referral process, delays in mitigation, no means for feedback from PHC/NRC | delays in mitigation, no means for feedback from NRC | difficulties with follow-up after discharge |

**Table 3. Summary of Challenges in integration**

## 8.6  Perceptions about SNEHA held by AWW's not using the application

*"Heard that she(AWW with SNEHA) is been given a better phone and sim card. The other day she showed me that and was telling it was easy to use. She marks the attendance in the phone. The best thing is… You just need to put the height and weight. You don't need to do any calculation or cross checks with growth charts. It automatically gives the color tab to show the SD. It is time saving. We are waiting to use this phone with app."*

From this study, it can be understood that the perceived usefulness (PU) [the degree to which they consider this application can improve their performance] and perceived ease of use (PEOU) [ the extent to which physical efforts required to use this application] seemed to be relatively high. For those who were not given this application, they considered it to be time saving and less of effort. The amount of time spent on finding the right book, drawing separators on it, maintaining and replicating it to consolidate for monthly reports and safeguarding them in their limited space was a great concern for AWWs. However, they feel that the chance of erring while cross checking with the growth chart also would be minimal when they use SNEHA app (PU). If they need to be using the new standards, then this mobile app could ease out their work as they were told that they need to input only height and weight (PEOU). Rest would be taken care by the device. To these AWWs who were using simple feature phones, android phone with all these applications was a matter of pride (A). However, it was also brought to the notice of the researcher that senior AWWs were slow on their input skills on mobile keypads. According to Davis Technology acceptance model (TAM), the actual system use is determined not only by PU and PEOU but together with the interplay of external variables in the context of functioning(Davis & Davis, 1989; Venkatesh, 2000). In this use case, effective and sustained use of SNEHA by any AWW is dependent on the external variables identified through ethnography and listed in Figure 2.

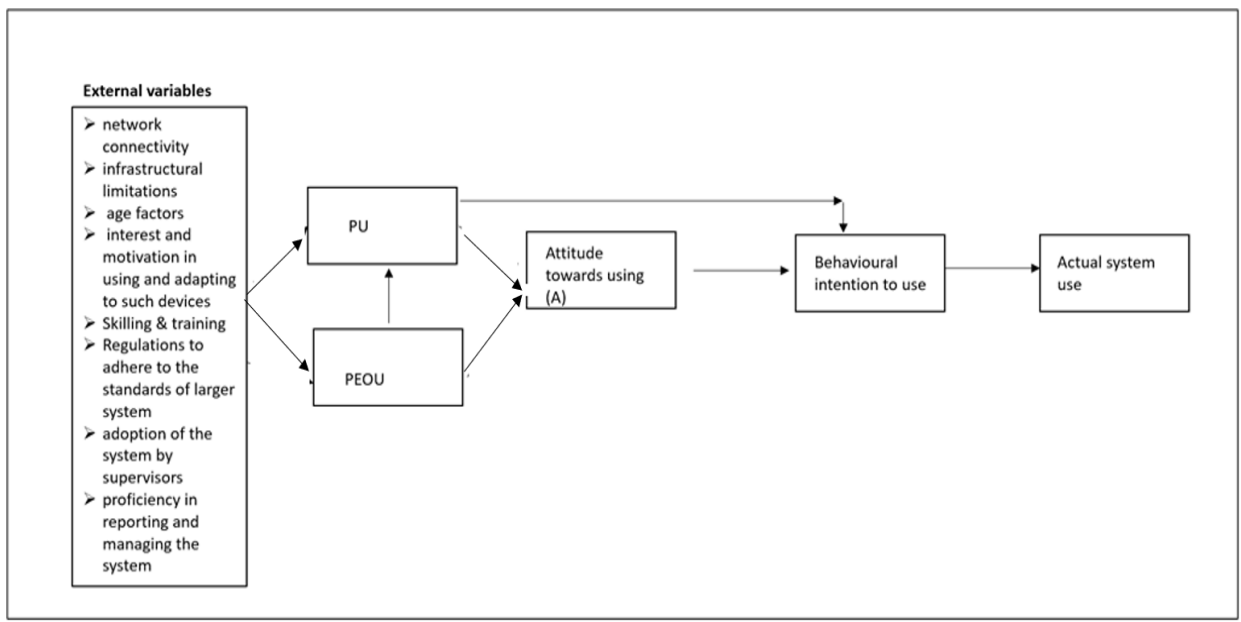

**Figure 2. Applying TAM model**

It can be inferred that the intention to use of such digital applications does not solely lie on usability and functionality features, rather the context in which they are to function also deserves special consideration in design, development and implementation. Nonetheless, such application in itself end up as yet another fragmented service.

## 9. DISCUSSION

We began with the premise that integrated care is the solution to overcome fragmented care and ICTs can be key enablers for integration of services. The use case of SNEHA in community-based SAM management throws light on the fact that digitization of an exact replica of paper-based system would not support integration of care unless the complexities in the context are not addressed. For an effective integration of care, there should be means of communication and information exchange of quality data between the nodes of care delivery, in this case between AWC, PHC and NRC. Ethnographic approach used in this study helped to understand the dynamics in natural settings and work practices of AWWs towards malnutrition management. Being the representatives of local community and the social proximity of the AWWs with their beneficiaries, they do exercise a significant role in understanding the community health needs, mediation of information exchange between health authorities and ensuring appropriate community based mitigatory measures. This finding is in agreement with another study on positive and impactful role of AWWs in articulating people's voice in the design and implementation of government policy and program(Prakash, 2015). The study has identified some of the contextual disconnects that results in 'fragmentation in care' for SAM management, even when the digital system is in place as listed in Table 3. Although, SNEHA anganwadi application has considerable achievements 'in silos' towards its objective of easing out daily activities of AWWs, its potential ability to address a complex problem like malnutrition can be harnessed by acknowledging the contextual dynamics like infrastructural deficiencies, resource constraints, ways and means to fix up process gaps, flexibility to accommodate the changing standards and appropriate trainings and skilling as required. Each of these contextual variables add to disconnectedness, thereby resultant fragmentation ensues. The qualitative data from this study is in alignment with the quantitative Quick Test Check Study (QTCS,2014) conducted by NITI Aayog, the results of which says that 24.3% of 10,92,877 AWCs studied presented with problems in manual record maintenance and 41% with inadequate infrastructures. While we acknowledge that perceived usefulness and perceived ease of use of SNEHA application appeared to be impressive, it is imperative to understand that the intention to use would be largely influenced by contextual dynamics. SNEHA application does have the potential to serve as a simple decision support tool for assessment of nutritional indicators, but, with wrong inputs provided, it ends up with misdiagnosis for SAM. Lack of right/ quality data and right connect of care delivery points can disrupt the continuum of care.

During the time of this study, the scope of SNEHA application was limited to Anganwadi only thereby limiting its functionality to mere screening purposes, as there was no information exchange with PHC or NRC for further referral and follow up. Hence, the service offered would still remain fragmented. Moreover, at the PHC and NRC levels, delays in mitigation in terms of functional (procuring medicines) or regulatory (getting sanctions for medical reimbursement) and lack of feedback mechanisms further adds to fragmentation in care. Disconnects between the doctor's advice during the convalescent period especially after discharge from NRC may cause the child to fall back to SAM category.

Therefore, ICTs when used in such interventions requiring continuum of care should be able to appreciate interactions between the recognized nodes of service delivery. Unless efforts are made to understand the perspectives of the interacting agents at each nodes of care delivery together with contextual dynamics in which such applications are to function, effective integration of care cannot be ensued. Such integrated systems must strengthen their resilience at three levels :a)Human system- in this study, training and skilling essential to AWWs to adapt to changing situations b) ICTs and Information systems- to be agile and accommodative of contextual requirements c) Organizations- to facilitate healthy interactions (Heeks & Ospina, 2018)

Limitation of the study**:** The research was carried out at a stage where only two domains of the SNEHA application (Child and Mother services) were active. Also, it was rolled out only to limited centres of the state of Karnataka. An overall evaluation of the application was not possible at this phase.

## 10.   CONCLUSION

Integrated care is suggested as a viable solution to overcome the problem of fragmentation in care, one of the shortcomings of complex health systems. ICTs can be a key enabler to such integrations by virtue of their ability to integrate and coordinate information across multiple systems. This qualitative study on AWWs and ICT in Nutritional management, has brought to light the contextual dynamics that can challenge effectiveness of ICT enabled integrated care. Mere digitization of paper-based systems cannot address the larger problem of fragmentation. The findings of the study contribute to the knowledge that unless efforts are made to understand the perspectives of the interacting agents at each nodes of care delivery together with contextual dynamics in which such ICTs are to function, effective integration of care cannot be ensued.